# Linear mixed model vs two-stage methods: Developing prognostic models of diabetic kidney disease progression

Brian Kwan[1,2], Lin Liu[1,2], David Strong[2], H. Irene Su[2,3], and Loki Natarajan[1,2]

[1]Division of Biostatistics and Bioinformatics, Department of Family Medicine and Public Health, University of California, San Diego, La Jolla, CA, USA;

[2]Moores Cancer Center, University of California, San Diego, La Jolla, CA, USA;

[3]Department of Obstetrics, Gynecology and Reproductive Sciences, University of California, San Diego, La Jolla, CA, USA.

**Corresponding author:**

Loki Natarajan, Division of Biostatistics, Department of Family Medicine and Public Health, University of California San Diego, 3855 Health Sciences Dr #0901, La Jolla, CA 92093, USA. Email: lnatarajan@ucsd.edu.

**Abstract**

Identifying prognostic factors for disease progression is a cornerstone of medical research. Repeated assessments of a marker outcome are often used to evaluate disease progression, and the primary research question is to identify factors associated with the longitudinal trajectory of this marker. Our work is motivated by diabetic kidney disease (DKD), where serial measures of estimated glomerular filtration rate (eGFR) are the longitudinal measure of kidney function, and there is notable interest in identifying factors, such as metabolites, that are prognostic for DKD progression. Linear mixed models (LMM) with serial marker outcomes (e.g., eGFR) are a standard approach for prognostic model development, namely by evaluating the time $\times$ prognostic factor (e.g., metabolite) interaction. However, two-stage methods that first estimate individual-specific eGFR slopes, and then use these as outcomes in a regression framework with metabolites as predictors are easy to interpret and implement for applied researchers. Herein, we compared the LMM and two-stage methods, in terms of bias and mean squared error via analytic methods and simulations, allowing for irregularly spaced measures and missingness. Our findings provide novel insights into when two-stage methods are suitable longitudinal prognostic modeling alternatives to the LMM. Notably, our findings generalize to other disease studies.

**Keywords**

Biomarker, disease progression, longitudinal study, mixed model, multilevel model, prognostic model

# 1 Introduction

Repeated longitudinal assessment of a marker of disease occurrence or progression is common in medical studies, e.g., serial measures of prostate specific antigen as a marker of prostate cancer, or repeated hemoglobin A1C for diabetes control (Lyons and Basu, 2012; O'Brien et al., 2011). Often, interest lies in identifying baseline factors associated with longitudinal trajectories of these markers, as these factors could provide early insights into actionable guidelines/treatments for the condition in question. Statistical methods for modeling these risk factor-longitudinal marker assessments is the focus of this article, with the specific research question motivated by our prior work in diabetic kidney disease (DKD) (Kwan et al., 2020).

Diabetes is a leading cause of kidney disease and patients with DKD are at high risk of morbidity, hospitalization, and overall mortality (*American Journal of Kidney Diseases*, 2018; Grams et al., 2017). Studies have shown that the human metabolome has considerable potential for characterizing patients with DKD versus healthy controls (Abbiss et al., 2019; Colhoun and Marcovecchio, 2018; Hirayama et al., 2012; Kalim and Rhee, 2017; Sharma et al., 2013; Zhang et al., 2015). By incorporating metabolomic analysis into statistical model development, we could construct prognostic models for early detection of patients at high risk of developing DKD, potentially leading to earlier and more targeted treatments. Estimated glomerular filtration rate (eGFR) is a clinically accepted method for measuring kidney function, with higher eGFR indicating better kidney function (Levey et al., 2009); slope of serial eGFR assessments, interpreted as annual eGFR change, are widely used to evaluate kidney disease progression. In our previous work (Kwan et al., 2020), we implemented a two-stage approach for identifying metabolomic predictors of DKD progression via, first estimating eGFR slope, and then using this slope as the outcome in a regression model with baseline metabolites as predictors. We used data collected from the Chronic Renal Insufficiency Cohort (CRIC) (Denker et al., 2015; Feldman, 2003; Lash et al., 2009), a racially and ethnically diverse group of adults aged 21 to 74 years with a broad spectrum of renal disease severity, one of the largest in the US, with comprehensive data on clinical and metabolite profiles. However, a more conventional and statistically accepted modeling approach is to fit a single linear mixed model with serial eGFR measures (outcomes) and evaluate the coefficient of the metabolite (biomarker) × time (year) interaction term, also interpreted as annual eGFR change. Nonetheless, two-stage methods offer the advantage of estimating individual slopes, which are by themselves of interest as a marker of disease progression, and can be readily implemented as outcomes in standard regression models by researchers, as evidenced by the plethora of research that uses eGFR slopes as outcomes in DKD research (Anderson et al., 2020; de Hauteclocque et al., 2014; Heinzel et al., 2018; Koye et al., 2018; Osonoi et al., 2020; Parsa et al., 2013). Given their widespread use by DKD researchers, in this paper, we aim to provide novel insights into when two-stage methods are suitable longitudinal prognostic modeling alternatives to the linear mixed model.

In prior statistical investigations, Sayers et al. (2017) conducted a simulation study comparing two-stage methods with individual slope as a predictor (i.e., independent variable) for a dependent outcome by examining the bias and coverage of the association between birth length, linear growth and later blood pressure under several study design scenarios. Our set-up is different in that the slopes are the dependent variable in our models, and we aim to evaluate a variety of two-stage approaches for assessing the prognostic value of a covariate for predicting this slope. In particular, using the framework of our previous work (Kwan et al., 2020), we will consider the baseline metabolite as the predictor for annual eGFR change (slope). In addition, expanding on the statistical approaches of Sayers et al. (2017), we compare via simulations the linear mixed effects model to our two-stage methods under an expanded set of study design scenarios that incorporate irregularly spaced time measures, and missing data and also analytically examine and compare bias and efficiency across methods. More specifically, in Section 2, we outline our statistical approaches which include a range of two-stage methods. In Section 3, we describe in detail our simulation process, study design scenarios, and comparison performance metrics for our statistical approaches. Section 4 showcases analytical derivations for the relationships between our statistical models. Section 5 presents the simulation results for our statistical approaches under our set of study design scenarios. Lastly, Section 6 discusses the overall findings, current limitations, and future directions for this work. We emphasize that although this paper is motivated by the metabolite-DKD context with the terms metabolite and eGFR serving as predictor and longitudinal outcome in the following sections, this work applies to any predictor-longitudinal disease modeling application.

## 2  Statistical Approaches

*2.1    Linear Mixed Model (LMM) Approach*

The linear mixed effects model (Fitzmaurice et al., 2011), ubiquitously used in longitudinal settings, incorporates fixed and random effects to model individual eGFR trajectories over time. Fixed effects are shared between all individuals and model the population mean eGFR trajectory. Random effects are unique to each individual and characterize individual eGFR profiles. Our model, which incorporated fixed effects for metabolite, time, and their interaction as well as random intercept and slope terms, was expressed as

$$y_{ij} = (\beta_0 + b_{0i} + \beta_1 * M_i) + (\beta_2 + b_{1i} + \beta_3 * M_i) * t_{ij} + \epsilon_{ij}$$

for individual $i$ and occasion $j$, where $y_{ij}$ is the eGFR response, $(\beta_0, \beta_1, \beta_2, \beta_3)$ are fixed effects and $(b_{0i}, b_{1i})$ are random effects, $M_i$ is individual $i$'s baseline metabolite value, $t_{ij}$ is time in years, and $\epsilon_{ij}$ is the within-individual error. We assume the random effects $(b_{0i}, b_{1i}) \sim N(\mathbf{0_2}, \mathbf{\Omega})$, where $\mathbf{\Omega} = \begin{pmatrix} \omega_0 & \omega_{01} \\ \omega_{10} & \omega_1 \end{pmatrix}$, are independent of both $M_i$ and $\epsilon_{ij}$. The within-individual error $\epsilon_{ij}$ is assumed to be normally distributed with mean zero and variance $\sigma^2$. As our investigation primarily focuses on the association between metabolite and annual rate of eGFR change, the $\beta_3$ metabolite × time interaction coefficient is our main effect of interest. The coefficient is interpreted as the population-averaged annual rate of eGFR change for a one-unit higher in metabolite value.

An advantage to using a linear mixed effects model is that it can incorporate incomplete and unbalanced longitudinal data among individuals. Therefore, we would be avoiding the bias of using complete-case analysis as well as not requiring an equal number of available eGFR measurements nor need these measurements be at a common set of occasions for each individual. A further, more extensive overview, of the method is given in Chapter 8 of Fitzmaurice et al. (2011) .

*2.2    Two-Stage Approaches*

Our two-stage methods model the association between metabolite and annual rate of eGFR change in two stages: (1st) estimate individual eGFR slopes and (2nd) regress eGFR slope on metabolite as the sole predictor. The first stage estimates individual eGFR slopes $C_i$ (for individual $i$) where the method of estimation varies between approaches. The second stage, similar for all approaches, fits a simple linear regression model with eGFR slope $C_i$, taken from the first stage, as the outcome on metabolite $M_i$.

$$\hat{C}_i = \alpha_0 + \alpha_1 * M_i + \epsilon_{i,SS}, \epsilon_{i,SS} \sim N(0, \sigma_{SS}^2)$$

The metabolite coefficient $\alpha_1$ is the association between metabolite and annual rate of eGFR change and is the population-averaged annual rate of eGFR change for a one-unit increase in metabolite value, which is interpreted similarly to the metabolite × time interaction coefficient $\beta_3$ from our linear mixed effects model.

### 2.2.1 Simple Approach

The simple approach to estimating eGFR slope is to take the difference between a subject's last and first observed eGFR measurements and divide by the elapsed time (years) between measurements.

$$\hat{C}_{i,\text{SIMPLE}} = \frac{(y_{iJ} - y_{i1})}{(t_{iJ} - t_{i1})}$$

This method contains a notable loss of in-between measurement information and calculates annual rate of eGFR change using only the latest and first observed eGFR measurements. Due to this loss of measurement information, $\hat{C}_{i,\text{SIMPLE}}$ will generally have greater variance than the true $C_i$.

### 2.2.2 Ordinary Lease Squares (OLS) Approach

We can also fit a simple linear regression model to the serial eGFR measures of an individual, with time as the predictor, to estimate the individual's eGFR slope.

$$y_{ij} = \gamma_{0i} + \gamma_{1i} * t_{ij} + \epsilon_{i,OLS}, \epsilon_{i,OLS} \sim N(0, \sigma_{OLS}^2)$$

The model parameters $\gamma_{0i}, \gamma_{1i}, \sigma_{OLS}^2$ are estimated by OLS. Let $\hat{\gamma}_{1i} = \hat{C}_{i,\text{OLS}}$ be the eGFR slope for individual $i$. By the Gauss-Markov theorem, $\hat{\gamma}_{1i}$ has the minimum sampling variance among the class of linear unbiased estimators for $\gamma_{1i}$. Since the individual slopes do not all provide equally precise information (differing number of individual eGFR measurements), $\hat{C}_{i,\text{OLS}}$ has greater variance than the true $C_i$. This approach requires fitting $I$ total models to estimate all of the individual eGFR slopes.

*2.2.3 Best Linear Unbiased Predictor (BLUP) Approach*

As opposed to fitting separate simple linear regression models for each individual, we can fit a single linear mixed-effects model to the longitudinal eGFR data of all individuals to estimate all of their eGFR slopes. Our model consisted of a fixed effect for time and random intercept and slope terms.

$$y_{ij} = (\eta_0 + u_{0i}) + (\eta_1 + u_{1i}) * t_{ij} + \epsilon_{ij,BLUP}$$

for individual $i$ and occasion $j$, where $y_{ij}$ is the eGFR response, $(\eta_0, \eta_1)$ are fixed effects and $(u_{0i}, u_{1i})$ are random effects, $t_{ij}$ is time in years, and $\epsilon_{ij,BLUP}$ is the within-individual error. We assume the random effects $(u_{0i}, u_{1i}) \sim N(\mathbf{0_2}, \mathbf{\Omega_{BLUP}})$, where $\mathbf{\Omega_{BLUP}} = \begin{pmatrix} \omega_{0,BLUP} & \omega_{01,BLUP} \\ \omega_{10,BLUP} & \omega_{1,BLUP} \end{pmatrix}$, are independent of $\epsilon_{ij,BLUP}$. The estimated random effects $(\hat{u}_{0i}, \hat{u}_{1i})$ are the best linear unbiased predictors (BLUPs) for the true $(u_{0i}, u_{1i})$. The within-individual error $\epsilon_{ij,BLUP}$ is assumed to be normally distributed with mean zero and variance $\sigma^2_{BLUP}$. Our estimated individual eGFR slopes are obtained by adding the estimated mean eGFR slope $\hat{\eta}_1$ to the estimated BLUP slopes $\hat{u}_{1i}$, i.e. let $(\hat{\eta}_1 + \hat{u}_{1i}) = \hat{C}_{i,\text{BLUP}}$.

Similarly, the model written in matrix notation is

$$\mathbf{Y} = \mathbf{X}\boldsymbol{\eta} + \mathbf{Z}\mathbf{u} + \boldsymbol{\epsilon}_{BLUP}$$

where $\mathbf{Y} = (\mathbf{Y_1}, \mathbf{Y_2}, \dots, \mathbf{Y_I})'$ s.t. $\mathbf{Y_i} = (y_{i1}, y_{i2}, \dots, y_{iJ})'$ and $Y$ is a vector of serial eGFR response values with length $I \times J$, $\mathbf{X} = (\mathbf{X_1}, \mathbf{X_2}, \dots, \mathbf{X_I})'$ s.t. $\mathbf{X_i}$ is the fixed effects design matrix for subject $i$ and $\dim(\mathbf{X}) = (I \times J) \times 2$, $\boldsymbol{\eta} = (\eta_0, \eta_1)'$, $\mathbf{Z} = \begin{pmatrix} \mathbf{Z_1} & & \\ & \ddots & \\ & & \mathbf{Z_I} \end{pmatrix}$ s.t. $\mathbf{Z_i}$ is the random effects design matrix for subject $i$ and $\dim(\mathbf{Z}) = (I \times J) \times (2 \times J)$, $\mathbf{u} = (\mathbf{u_1}, \mathbf{u_2}, \dots, \mathbf{u_I})'$ s.t. $u_i = (u_{0i}, u_{1i})'$ and $u$ is a vector of random effects with length $2 \times J$, and $\boldsymbol{\epsilon}_{BLUP} = (\boldsymbol{\epsilon}_{1,BLUP}, \boldsymbol{\epsilon}_{2,BLUP}, \dots, \boldsymbol{\epsilon}_{I,BLUP})'$ s.t. $\boldsymbol{\epsilon}_{i,BLUP} = (\epsilon_{i1,BLUP}, \epsilon_{12,BLUP}, \dots, \epsilon_{iJ,BLUP})'$ and $\boldsymbol{\epsilon}_{BLUP}$ is a vector of eGFR measurement errors with length $I \times J$. For our setup, we assume $\mathbf{X_i} = \mathbf{Z_i}$ since our model consisted of only a fixed effect for time, while having both random intercept and slope terms. We assume $u \sim N(0, \mathbf{G_{BLUP}})$, where $\mathbf{G_{BLUP}} = \begin{pmatrix} \Omega_{BLUP} & & \\ & \ddots & \\ & & \Omega_{BLUP} \end{pmatrix}$, and $\epsilon_{BLUP} \sim N(0, \sigma^2_{BLUP}I)$ are independent of each other. We note that the BLUPs $\hat{u}$ are a weighted average of the population- and individual-level counterparts, and hence will have lower variance than the true values. We discuss a way to address this in the next section.

*2.2.4 Inflated Approach*

To address the under-estimation of variances of the BLUP random effects in comparison to its restricted maximum likelihood (REML) estimation for the covariance matrix $\mathbf{G_{BLUP}}$, Carpenter et al. (2003) transformed (re-inflated) the random effects so that their crude covariance matrix is more equivalent to $\mathbf{G_{BLUP}}$. The re-inflated random effects are then added to the estimated fixed effects intercept $\hat{\eta}_0$ and slope $\hat{\eta}_1$ to give the estimated eGFR baseline value and slope, respectively, for each individual.

Here we briefly state the analytic steps as described by Sayers et al. (2017). The re-inflation process involves multiplying our estimated random effects matrix by an upper triangular matrix of equal order. Hence, we require finding a transformation $\mathbf{A}$ such that

$$\widehat{\mathbf{U}}^* = \widehat{\mathbf{U}}\mathbf{A}$$

where $\widehat{\mathbf{U}}^*$ is the matrix of the inflated random effects and $\widehat{\mathbf{U}}$ is the matrix of our originally estimated random effects, both with $I$ rows and 2 columns. The matrix $\mathbf{A}$ is formed using the lower triangular Cholesky decompositions of the empirical covariance matrix of the estimated random effects as well as its corresponding REML covariance matrix. The empirical covariance matrix is calculated as

$$\mathbf{S} = \widehat{\mathbf{U}}^T\widehat{\mathbf{U}} / N$$

and the REML covariance matrix as

$$\mathbf{R} = \widehat{\Omega}_{BLUP}$$

and $\mathbf{S}$ and $\mathbf{R}$ written in terms of their lower triangular Cholesky decompositions are

$$S = L_S L_S^T$$

$$R = L_R L_R^T$$

Finally, $A$, an upper triangular matrix can be calculated as

$$A = \left(L_R L_S^{-1}\right)^T$$

The transformed (re-inflated) random effects $\widehat{U}^*$ now have covariance matrix equivalent to that of the model estimate $\widehat{\Omega}_{\text{BLUP}}$.

## 3  Simulation study design

We compare our statistical approaches, i.e., linear mixed model vs two-stage methods, under various study design scenarios. Since the linear mixed model is the more conventional method for modeling disease progression, it served as the data generating model for our simulated study. Our model consisted of fixed effects for metabolite, time (i.e., year of follow-up) and their interaction as well as random intercept and slope terms. The number of individuals in our study ($I$) was set to 200, representing a medium sized study cohort, and individuals had eGFR measurements biennially from 0 to 10 years of follow-up ($J = 6$). For scenarios with missing eGFR data, value $J$ will vary by individual ($J_i \leq 6$). Regardless, we use $J$ for our model notation. We compared the bias and efficiency of our 4 two-stage statistical modeling approaches across study design scenarios based on differing (1) choice of spacing between eGFR measures (regularly vs irregularly spaced), (2) amount of missing completely at random (MCAR) eGFR data (complete, 20%, 50%, 80%), and (3) standard deviation (SD) value for the metabolite, random intercept, random slope, and measurement error as well as the correlation value between the random effects in our data generating model. Our chosen values are as follows

(a) $\sigma_M$ (Metabolite) = (0.79), 2, 7, 10, 15, 20

(b) $\omega_0$ (Random Intercept) = 0.5, 1, 4, 7, (9.87), 12, 16

(c) $\omega_1$ (Random Slope) = 0.5, 1, (2.27), 4, 7, 10

(d) $\sigma_{err}$ (Error) = 0.5, 1, 3, (5.87), 8, 10, 15

(e) $\rho_\omega$ (Random Effects Corr.) = -1, -0.75, -0.5, -0.25, 0, (0.159), 0.25, 0.5, 0.75, 1

When varying a particular simulation parameter (e.g., metabolite SD), the values for the other parameters (i.e., random intercept SD, random slope SD, error SD, and random effects correlation) were held fixed at the value shown in parentheses for that parameter. The parameter values in parentheses for the varying parameters were selected for our data generating model as they were the numerical estimates from the linear mixed model fitted to the analytic cohort of the Chronic Renal Insufficiency (CRIC) Study from our previous work (Kwan et al., 2020). In addition, again following from our previous work, the fixed effect parameters and the mean of the metabolite were fixed for all simulations as follows:

(a) $\beta_0 = -24.22$

(b) $\beta_1 = 4.34$

(c) $\beta_2 = -5.13$

(d) $\beta_3 = 0.22$

(e) $\mu_M$ (Metabolite Mean) = 14.8

In total, we studied 48 different scenarios and generated $D = 1000$ replications for each of them to assess the performance of our statistical approaches. We compared the performance of the different methods in estimating the association between annual rate of eGFR change and metabolite for the linear mixed model $(\hat{\beta}_3)$ versus two-stage methods $(\hat{\alpha}_1)$ by examining the (relative) bias and efficiency, i.e. standard deviation (SD), standard error (SE), and root mean square error (MSE), across methods. The bias, relative bias, SD, and SE are defined as the following:

$$\text{Bias} = \left(\frac{1}{D} \sum_{d=1}^{D} \hat{\alpha}_{1,d}\right) - \beta_3$$

$$\text{Rel. Bias (\%)} = \frac{\text{Bias}}{\beta_3} \times 100$$

$$\text{Standard Deviation} = \sqrt{\frac{1}{D-1} \sum_{d=1}^{D} \left(\hat{\alpha}_{1,d} - \frac{1}{D} \sum_{d=1}^{D} \hat{\alpha}_{1,d}\right)^2}$$

$$\text{Standard Error} = \frac{1}{D} \sum_{d=1}^{D} \text{SE}(\hat{\alpha}_{1,d})$$

where $D$ is the total number of replications and root MSE is calculated as $\sqrt{\text{Bias}^2 + \text{SD}^2}$. The notation here uses our estimated association from the two-stage models ($\hat{\alpha}_{1,d}$), so calculating these statistics of interest for the linear mixed model would require replacing $\hat{\alpha}_{1,d}$ with $\hat{\beta}_{3,d}$.

Simulation study design and statistical analysis was conducted using the R (version 3.6.1) programming environment (R Core Team, 2019).

# 4 Analytical Relationships Between Statistical Models

*4.1 Unbiased association for the Simple and OLS methods*

We prove analytically that the Simple and OLS methods have unbiased association for the study design scenario with regularly spaced measures and complete data. In particular, our general second-stage model was

$$\hat{C}_i = \alpha_0 + \alpha_1 * M_i + \epsilon_{i,\text{SS}}, \epsilon_{i,\text{SS}} \sim N(0, \sigma^2_{\text{SS}})$$

and we show that $E(\hat{\alpha}_1) = \beta_3$ with $\hat{C}_{i,\text{Simple}}$ or $\hat{C}_{i,\text{OLS}}$ as the outcome. The coefficient $\alpha_1$ has estimate

$$\hat{\alpha}_1 = \frac{\sum_{i=1}^{I}(M_i - \bar{M})\hat{C}_i}{\sum_{i=1}^{I}(M_i - \bar{M})^2}$$

where $\bar{M} = \frac{1}{I}\sum_{i=1}^{I} M_i$. We can rewrite $\hat{C}_{i,\text{Simple}}$ based on our data generating model and obtain

$$\hat{C}_{i,\text{SIMPLE}} = (\beta_2 + \beta_3 * M_i + b_{1i}) + \frac{(\epsilon_{iJ} - \epsilon_{i1})}{(t_{iJ} - t_{i1})}$$

Our first-stage model in the OLS approach was

$$y_{ij} = \gamma_{0i} + \gamma_{1i} * t_{ij} + \epsilon_{i,\text{OLS}}, \epsilon_{i,\text{OLS}} \sim N(0, \sigma^2_{OLS})$$

and we let $\hat{\gamma}_{1i} = \hat{C}_{i,\text{OLS}}$ be the eGFR slope for individual $i$ such that

$$\hat{C}_{i,\text{OLS}} = \frac{\sum_{j=1}^{J}(t_{ij} - \bar{t}_i)(y_{ij} - \bar{y}_i)}{\sum_{j=1}^{J}(t_{ij} - \bar{t}_i)^2}$$

where $\bar{t}_i = \frac{1}{J}\sum_{j=1}^{J} t_{ij}$ and $\bar{y}_i = \frac{1}{J}\sum_{j=1}^{J} y_{ij}$. Similarly, we can write $\hat{C}_{i,\text{OLS}}$ based on our data generating model and obtain

$$\hat{C}_{i,\text{OLS}} = (\beta_2 + \beta_3 * M_i + b_{1i}) + \frac{\sum_{j=1}^{J}(t_{ij} - \bar{t}_i)(\epsilon_{ij} - \bar{\epsilon}_i)}{\sum_{j=1}^{J}(t_{ij} - \bar{t}_i)^2}$$

where $\bar{\epsilon}_i = \frac{1}{J}\sum_{j=1}^{J}\epsilon_{ij}$. We can write $\hat{C}_{i,\text{OLS}}$ as a function of $\hat{C}_{i,\text{SIMPLE}}$

$$\hat{C}_{i,\text{OLS}} = \hat{C}_{i,\text{SIMPLE}} - \frac{(\epsilon_{iJ} - \epsilon_{i1})}{(t_{iJ} - t_{i1})} + \frac{\sum_{j=1}^{J}(t_{ij} - \bar{t}_i)(\epsilon_{ij} - \bar{\epsilon}_i)}{\sum_{j=1}^{J}(t_{ij} - \bar{t}_i)^2}$$

and if $J = 2$, then $\hat{C}_{i,\text{OLS}} = \hat{C}_{i,\text{SIMPLE}}$.

Defining $\hat{C}_{i,SIMPLE}$ based on our data generating model and having it as the outcome for the second-stage model, the estimated association $\hat{\alpha}_1$ is

$$\hat{\alpha}_1 = \frac{\sum_{i=1}^{I}\left\{(M_i - \bar{M})\left[\beta_2 + \beta_3 * M_i + b_{1i} + \frac{(\epsilon_{iJ} - \epsilon_{i1})}{(t_{iJ} - t_{i1})}\right]\right\}}{\sum_{i=1}^{I}(M_i - \bar{M})^2}$$

Taking the expected value, we have

$$E(\hat{\alpha}_1) = \beta_3 * \frac{\sum_{i=1}^{I}(M_i - \bar{M})M_i}{\sum_{i=1}^{I}(M_i - \bar{M})^2}$$

and by simplifying we have $E(\hat{\alpha}_1) = \beta_3$ and conclude that using the Simple slopes for our Two-Stage method give an unbiased association between annual rate of eGFR change and metabolite.

Similarly, defining $\hat{C}_{i,\text{OLS}}$ based on our data generating model and having it as the outcome for the second-stage model, the estimated association $\hat{\alpha}_1$ is

$$\hat{\alpha}_1 = \frac{\sum_{i=1}^{I}\left\{(M_i - \bar{M})\left[\beta_2 + \beta_3 * M_i + b_{1i} + \frac{\sum_{j=1}^{J}(t_{ij} - \bar{t}_i)(\epsilon_{ij} - \bar{\epsilon}_i)}{\sum_{j=1}^{J}(t_{ij} - \bar{t}_i)^2}\right]\right\}}{\sum_{i=1}^{I}(M_i - \bar{M})^2}$$

Taking the expected value, we have $E(\hat{\alpha}_1) = \beta_3$ and conclude that using the OLS slopes for our Two-Stage method also give an unbiased association between annual rate of eGFR change and metabolite.

*4.2    Correction of association bias for the BLUP method*

In contrast, our BLUP method will contain noticeable bias for the association between annual rate of eGFR change and metabolite, assuming that the $\beta_3$ metabolite × time interaction coefficient is the true association. We first derive the bias analytically, and then show how to correct for this bias by a transformation matrix for our estimated random effects (intercept & slope). Like before, we assume the study design scenario with regularly spaced measures and complete data.

In order to derive the parameters of interest, recall that our general second-stage model was

$$\hat{C}_i = \alpha_0 + \alpha_1 * M_i + \epsilon_{i,SS}, \epsilon_{i,SS} \sim N(0, \sigma_{SS}^2)$$

and our goal is to estimate $\hat{\alpha}_1$ with $\hat{C}_{i,BLUP}$ as the outcome. As noted in Section 2.2.3, individual eGFR (BLUP) slopes are obtained by adding the estimated mean eGFR slope $\hat{\eta}_1$ to the estimated BLUP slopes $\hat{u}_{1i}$, i.e. let $(\hat{\eta}_1 + \hat{u}_{1i}) = \hat{C}_{i,BLUP}$. Using standard mixed model theory we know that the $\widehat{U}$ matrix of our estimated (centered) random effects (intercept & slope) is indexed by $I$ rows and 2 columns, and can be estimated as $\widehat{U} = \mathbf{G_{BLUP}}Z^T V^{-1}(Y - X\widehat{\beta})$, where $\widehat{\beta} = (X^T V^{-1} X)^{-1} X^T V^{-1}$ and $V = Z\mathbf{G_{BLUP}}Z^T + \sigma_{BLUP}^2 I_N$ (Fitzmaurice et al., 2011). Having estimated the BLUP slopes, the second step of our two-stage method is to regress this BLUP slope on the metabolite predictor, i.e., to estimate $\hat{\alpha}_1$. However, although our main focus is on the BLUP slope, for ease of theoretical development, we will use matrix notation, and consider the regression problem $E(\widehat{U}| M)$, i.e., include random intercept and slope, and evaluate the expected value of our estimated random effects conditioned on the metabolite predictor. Using algebraic manipulations we see that:

$$E(\widehat{U}| M) = E(\mathbf{G_{BLUP}}Z^T V^{-1}(Y - X\widehat{\beta})| M)$$

$$= E(\mathbf{G_{BLUP}}Z^T V^{-1} HZU)| M) \text{ where } H = I_N - X(X^T V^{-1} X)^{-1} X^T V^{-1}$$

$$= \mathbf{G_{BLUP}}Z^T V^{-1} HZ * E(U|M)$$

We can see that regressing the estimated random effects results in a multiplicative bias matrix of $\mathbf{G_{BLUP}}Z^T V^{-1} HZ$ on the true random effects $U$. Thus except in the unlikely scenario that this bias matrix is the identity, use of BLUP slopes will result in biased estimates. We could correct for this bias by taking the inverse of this bias as a transformation matrix for our estimated random effects and multiply it to both sides.

$$(\mathbf{G_{BLUP}}Z^T V^{-1} HZ)^{-1} E(\widehat{U}|M) = E(U|M)$$

The recalculated random effects $(\mathbf{G_{BLUP}}Z^T V^{-1} HZ)^{-1}\widehat{U}$ will yield both transformed intercepts and slopes for individuals, which when the slope is used as the outcome for the second-stage model gives an unbiased association between annual rate of eGFR change and metabolite, assuming that the $\beta_3$ metabolite × time interaction coefficient is the true association. We apply this correction for our BLUP method in the simulation study; however, calculating the inverse of $\mathbf{G_{BLUP}}Z^T V^{-1} HZ$ proved to be unfeasible in study design scenarios with irregularly spaced time measures or MCAR data.

## 5 Simulation Results

We compared the bias and efficiency of the linear mixed model to our two-stage methods under our simulation study design, with the linear mixed model as the data generating model. We organized our simulation results based on varying a certain parameter in our data generating model. The text, table, and figures elaborate on the results for Complete Data and MCAR 50% and we describe the results in the text for MCAR 20% and 80% in relation to Complete Data and MCAR 50%.

### 5.1 No Varying Parameters

Table 1 shows the results. There were similar results for having regularly spaced and irregular spaced time measures in Complete Data. The LMM, Simple, and OLS methods have negligible bias supporting our analytic solution (Section 4.1) of the Simple and OLS methods having unbiased association. There is notable upward and downward bias for the BLUP and Inflated methods, respectively. However, after correcting for the bias in our BLUP slopes from our proposed analytic solution (Section 4.2), the BLUP had minimal bias (0.004) equal to that of the LMM, Simple, and OLS methods.

The BLUP and Inflated methods displayed overall greater efficiency than the other methods in having lower SD, SE, and root MSE. These results also hold for regularly spaced time measures in MCAR 50%. However, for irregularly spaced time measures in MCAR 50%, the Simple and OLS methods have overwhelmingly large bias and worse efficiency while the BLUP and Inflated performed similarly as in the aforementioned scenarios. For Complete Data scenarios or regularly spaced assessments, when comparing bias and efficiency, all of the two-stage methods are well-suited for modeling the association between eGFR slope and metabolite, with a notable bias-variance trade off in the BLUP and Inflated method. However, for irregularly spaced time measures with 50% missingness under a MCAR mechanism, we do not recommend using the Simple and OLS methods, as these displayed large bias and root MSE.

Results for MCAR 20%, in both the regularly and irregularly spaced cases, and MCAR 80%, in just the regularly spaced case, were similar to those of Complete Data; results for the irregularly spaced case for MCAR 80% were similar to the same case for MCAR 50%.

## 5.2 Vary Metabolite SD

Figure 1 shows the results. Similar results hold across the regularly and irregularly spaced time measures for Complete Data and the regularly spaced case for MCAR 50%. The LMM, Simple, and OLS have low relative bias across the spectrum of metabolite SD values, while the BLUP has notable upwards relative bias and the Inflated having downwards relative bias for lower metabolite SD values. However, with increasing metabolite SD values, the relative bias shrinks toward zero for both the BLUP and Inflated methods. Even with the bias in the BLUP and Inflated methods, they remain competitive to the LMM, Simple, and OLS methods in root MSE due to their lower SD values for their estimated metabolite associations (Supplementary Figure 1). In contrast with the scenario of irregularly spaced measures in MCAR 50%, the Simple and OLS methods have overwhelmingly large relative bias and SD, particularly for lower metabolite SD values which is reflected in their root MSE performance.

Results for the regularly spaced case for MCAR 20% and 80% were similar to those of the regularly and irregularly spaced cases for Complete Data and the regularly spaced case for MCAR 50%. The Simple and OLS in the irregularly spaced case for MCAR 20% had larger SD, SE, and root MSE across metabolite SD values than when they were in the aforementioned scenarios, while the irregularly spaced case for MCAR 80% shared similar results to the scenario of irregularly spaced measures in MCAR 50%.

5.3   *Vary Random Slope SD*

Figure 2 shows the results. For regularly spaced time measures in Complete Data and MCAR 50%, the LMM, Simple, and OLS methods have lower relative bias across all random slope SD values. The BLUP and Inflated methods both start off with notable downward relative bias with the BLUP spiking and then leveling off on a consistent upwards relative bias trend and the Inflated spiking before proceeding on a decreasing trend. The root MSE was similar across all methods. Similar results were observed for irregularly spaced time measures in Complete Data except for a noticeable spike of increasing relative bias for all methods at our last random slope SD value (15), where the Inflated method had the lowest relative bias. However, for irregularly space time measures in MCAR 50%, the relative bias is noticeably worse for the Simple and OLS methods with the magnitude > 100% for lower random slope SD values. The SD values were many times larger for the Simple and OLS methods (Supplementary Figure 2) resulting in their consistently higher root MSE than the other methods.

Results for the regularly spaced case for MCAR 20% were similar to those of the irregularly spaced case for Complete Data. Both the regularly spaced case for MCAR 80% and irregularly spaced case for MCAR 20% shared similar results to the regularly spaced case for MCAR 50%, but with larger SD, SE, and root MSE across random slope SD values. Finally, the irregularly spaced case for MCAR 80% shared similar results to the scenario of irregularly spaced measures in MCAR 50%.

5.4   *Vary Correlation between Random Intercept and Slope*

Figure 3 shows the results. For regularly and irregularly spaced time measures in Complete Data, the LMM, Simple, and OLS methods have small relative bias across correlation values. The Inflated method performs similarly except that for perfect negative or positive correction there is notable upward relative bias.

However, the BLUP method has a generally increasing trend in relative bias with greater correlation values. All methods performed competitively in root MSE. Similar results hold for regularly spaced time measures in MCAR 50% but there is more noticeable separation in root MSE performance with higher values for the Simple and OLS methods for non-perfect correlation values. This is attributed to the noticeable separation in SD performance across correlation values with the Simple and OLS displaying greater variability (Supplementary Figure 3). In contrast, for irregularly spaced time measures in MCAR 50%, the Simple and OLS methods have overwhelmingly large relative bias on much worse efficiency leading to both methods have much larger root MSE across correlation values.

The regularly spaced case for MCAR 20% had similar results to having regularly and irregularly spaced time measures in Complete Data. Similar results were true for the irregularly spaced case for MCAR 20% and the regularly spaced case for MCAR 80%, but with the Simple and OLS having larger SD, SE, and root MSE than the other methods across all correlation values. Finally, the irregularly spaced case for MCAR 80% shared similar results to the scenario of irregularly spaced measures in MCAR 50%.

## 5.5 Vary Random Intercept SD

The results for varying the random intercept SD are very similar to results from varying metabolite SD with a few exceptions. For regularly and irregularly spaced time measures for Complete Data and just the regularly spaced case for MCAR 50%, the BLUP method has slightly larger root MSE than the other methods for lower random intercept SD values (Supplementary Figure 4a). For irregularly spaced time measures in MCAR 50%, the Simple and OLS methods also have noticeably larger root MSE from having larger relative bias and SD than the other methods (Supplementary Figure 4b).

Results for MCAR 80% are similar to that of MCAR 50%. Furthermore, results of MCAR 20% are similar to that of Complete Data, except that the Simple and OLS methods had overall larger SD, SE, and root MSE than the other methods for the irregularly spaced case compares to the same case for Complete Data.

## 5.6 Vary Error SD

For regularly and irregularly spaced time measures for Complete Data and just the regularly spaced case for MCAR 50%, the LMM, Simple, and OLS methods have low relative bias across all error SD values, while the BLUP and Inflated methods have growing upward and downward relative bias for increasing error SD values, respectively (Supplementary Figure 5a). The BLUP and Inflated methods have lower SD values across all error SD values and thus performed similarly or better in root MSE than the other methods (Supplementary Figure 5b). For irregularly spaced time measures in MCAR 50%, there was both lack of consistent direction in and relatively larger relative bias for the Simple and OLS methods, particularly for larger error SD values. In addition, the SD and root MSE for both the Simple and OLS methods displayed a monotonically increasing trend for increasing error SD values.

Results for MCAR 80% are similar to that of MCAR 50%. Furthermore, results of MCAR 20% are similar to that of Complete Data, except that the Simple and OLS methods had larger relative bias, SD, SE, and root MSE than the other methods for larger error SD values in the irregularly spaced case of MCAR 20%.

## 6  Discussion

We have uncovered study design scenarios where two-stage methods are well-suited modeling alternatives to the linear mixed model by comparing the association between metabolite and annual eGFR change. For regularly and irregular spaced time measures in Complete Data and just regularly spaced time measures in MCAR 50%, the Simple and OLS methods have lower bias than the BLUP and Inflated methods. However, we have shown that the BLUP method can correct bias and both the BLUP and Inflated methods have greater efficiency with lower SD, SE, and root MSE. This provides credence to our previous work (Kwan et al., 2020) in using the BLUP approach to estimate eGFR slopes. Also, with regularly spaced or complete data, we saw that increasing the SD of metabolite or random intercept is associated with a decreasing trend in the bias for the BLUP and Inflated methods. Furthermore, with regularly spaced or complete data across random slope SD, random effects correlation, and error SD values, the Simple and OLS methods performed much more favorably in bias with the trade-off of slightly worse efficiency compared to the BLUP and Inflated methods. Thus in these scenarios the choice of optimal method will be dictated by the goals of the analysis, namely whether to minimize overall prediction error vs unbiased estimation of associations. Most importantly, throughout our simulation study when varying parameters, the Simple and OLS methods performed noticeably worse in statistical performance with irregularly spaced time measures and MCAR 50% data, scenarios that are not uncommon in observation studies, and so we do not recommend either of these methods when both data criterion are met.

We acknowledge limitations of our work. First, our linear mixed model and two-stage methods assume eGFR has a linear rate of change. Statistical models that account for nonlinear trajectories should be considered; however, despite the rigid linearity assumption, eGFR slopes are an established, clinically useful, and commonly used measure of diabetic kidney disease progression (Anderson et al., 2020; de Hauteclocque et al., 2014; Heinzel et al., 2018; Koye et al., 2018; Kwan et al., 2020; Osonoi et al., 2020; Parsa et al., 2013). Second, the chosen number of subjects N=200 for our simulation study design compares our statistical approaches under a medium sized cohort and further analysis with smaller and larger N could provide additional guidance on optimal choice of methods for small and large sized cohorts, respectively. Third, we have only compared our statistical approaches under a single missing data mechanism, MCAR. More complex missing data mechanisms could also arise such as data that is missing at random (MAR) or missing not at random (MNAR). Further investigation of these topics would require additional simulation scenarios and assumptions; we aim to investigate this in future studies.

Our work has elucidated the choice between the linear mixed model vs two-stage methods for predicting possible patient future disease progression based on their clinic entry biomarker data under various study design scenarios. Although the linear mixed model is an optimal approach, there were numerous scenarios where at least one two-stage method was a suitable modeling alternative to mixed models, which opens the doors for clinicians to implement standard statistical methods using slope outcomes. Similar to Sayers et al. (2017), we examined a single continuous biomarker predictor and additional studies looking into adjusting for key clinical risk factors and confounders that could further improve prognostication of disease progression, e.g., baseline eGFR (Grams et al., 2018), will further illuminate our modeling options across various study design scenarios. However, including covariates would require assumptions on joint covariate distributions and additional simulations, and we do not pursue this further here. Importantly, in our single marker setting, we were able to analytically calculate bias (or lack thereof) for the proposed two-stage methods, and propose a method to mathematically correct this bias.

In summary, in this work via simulations and analytic calculations, we evaluated two-stage methods for estimating marker-DKD progression associations in a longitudinal setting. We examined a range of realistic study designs commonly encountered in medical research (e.g., irregularly spaced measures, missing data), and identified scenarios where two-stage models performed competitively. Of note, for many disease settings (e.g., eGFR trajectory and kidney disease, prostate-specific-antigen change and prostate cancer, rate of decline in FEV and chronic obstructive pulmonary disease) (Celli et al., 2008; Li et al., 2012; O'Brien et al., 2011), the rate of change (i.e. slope) of the biomarker is of interest in its own right as a marker of disease, and thus is often an outcome of interest. Thus our findings are easily generalizable to other disease prognostic modeling studies.

**Acknowledgements**

This material is based upon work supported by the National Science Foundation Graduate Research Fellowship Program under Grant No. DGE-1650112. Any opinions, findings, and conclusions or recommendations expressed in this material are those of the author(s) and do not necessarily reflect the views of the National Science Foundation.

**Disclosure of financial and other interests**

The authors declare that there are no conflicts of interest.

**Funding**

BK was supported by the National Science Foundation Graduate Research Fellowship Program [grant number DGE-1650112]. LN was partially supported by the National Institute of Diabetes and Digestive and Kidney Diseases [grant number 1R01DK110541-01A1].

Table 1: Comparison of simulation results (D=1000, N=200) for the estimated association between annual rate of eGFR change and metabolite. True association $\beta_3 = 0.223$.

(a) Complete Data

| Statistic | LMM | Simple | OLS | BLUP | Inflated |
|---|---|---|---|---|---|
| Bias | 0.004 (-0.003) | 0.004 (-0.003) | 0.004 (-0.003) | 0.071 (0.073) | -0.01 (-0.016) |
| Rel. Bias (%) | 1.66 (-1.202) | 1.66 (-1.507) | 1.66 (-1.175) | 31.959 (32.666) | -4.279 (-7.387) |
| SD | 0.212 (0.217) | 0.216 (0.225) | 0.212 (0.219) | 0.194 (0.195) | 0.203 (0.205) |
| SE | 0.214 (0.217) | 0.218 (0.223) | 0.214 (0.217) | 0.196 (0.193) | 0.205 (0.205) |
| Root MSE | 0.212 (0.217) | 0.216 (0.225) | 0.212 (0.219) | 0.207 (0.209) | 0.203 (0.206) |

(b) MCAR 50%

| Statistic | LMM | Simple | OLS | BLUP | Inflated |
|---|---|---|---|---|---|
| Bias | -0.003 (0.003) | -0.008 (7.631) | -0.007 (7.64) | 0.136 (0.141) | -0.034 (-0.029) |
| Rel. Bias (%) | -1.43 (1.127) | -3.398 (3421.86) | -3.187 (3426.183) | 60.998 (63.097) | -15.124 (-13.145) |
| SD | 0.241 (0.259) | 0.275 (243.471) | 0.274 (243.472) | 0.189 (0.19) | 0.208 (0.216) |
| SE | 0.24 (0.249) | 0.272 (20.169) | 0.271 (20.17) | 0.181 (0.177) | 0.206 (0.206) |
| Root MSE | 0.242 (0.259) | 0.275 (243.591) | 0.274 (243.591) | 0.233 (0.237) | 0.21 (0.218) |

[†]Results displayed as: Regularly Spaced case (Irregularly Spaced case).
LMM, Linear Mixed Model; OLS, Ordinary Least Squares; BLUP, Best Linear Unbiased Predictor; SD, Standard Deviation; SE, Standard Error; MSE, Mean Squared Error.

Bias $= \left(\frac{1}{D} \sum_{d=1}^{D} \hat{\alpha}_{1,d}\right) - \beta_3$; Rel. Bias (%) $= \frac{\text{Bias}}{\beta_3} \times 100$;

Standard deviation $= \sqrt{\frac{1}{D-1} \sum_{d=1}^{D} \left(\hat{\alpha}_{1,d} - \frac{1}{D} \sum_{d=1}^{D} \hat{\alpha}_{1,d}\right)^2}$;

Standard Error $= \frac{1}{D} \sum_{d=1}^{D} \text{SE}(\hat{\alpha}_{1,d})$;

Root MSE $= \sqrt{\text{Bias}^2 + \text{SD}^2}$

Main Figure 1 (Rel. Bias, Root MSE for vary metabolite SD)
Performance in relative bias (%) and root MSE of our methods in estimating the association between annual rate of eGFR change and metabolite for the linear mixed model ($\hat{\beta}_3$) versus two-stage methods($\hat{\alpha}_1$) as a function of metabolite SD for the regularly and irregularly spaced cases of Complete Data and MCAR 50%.

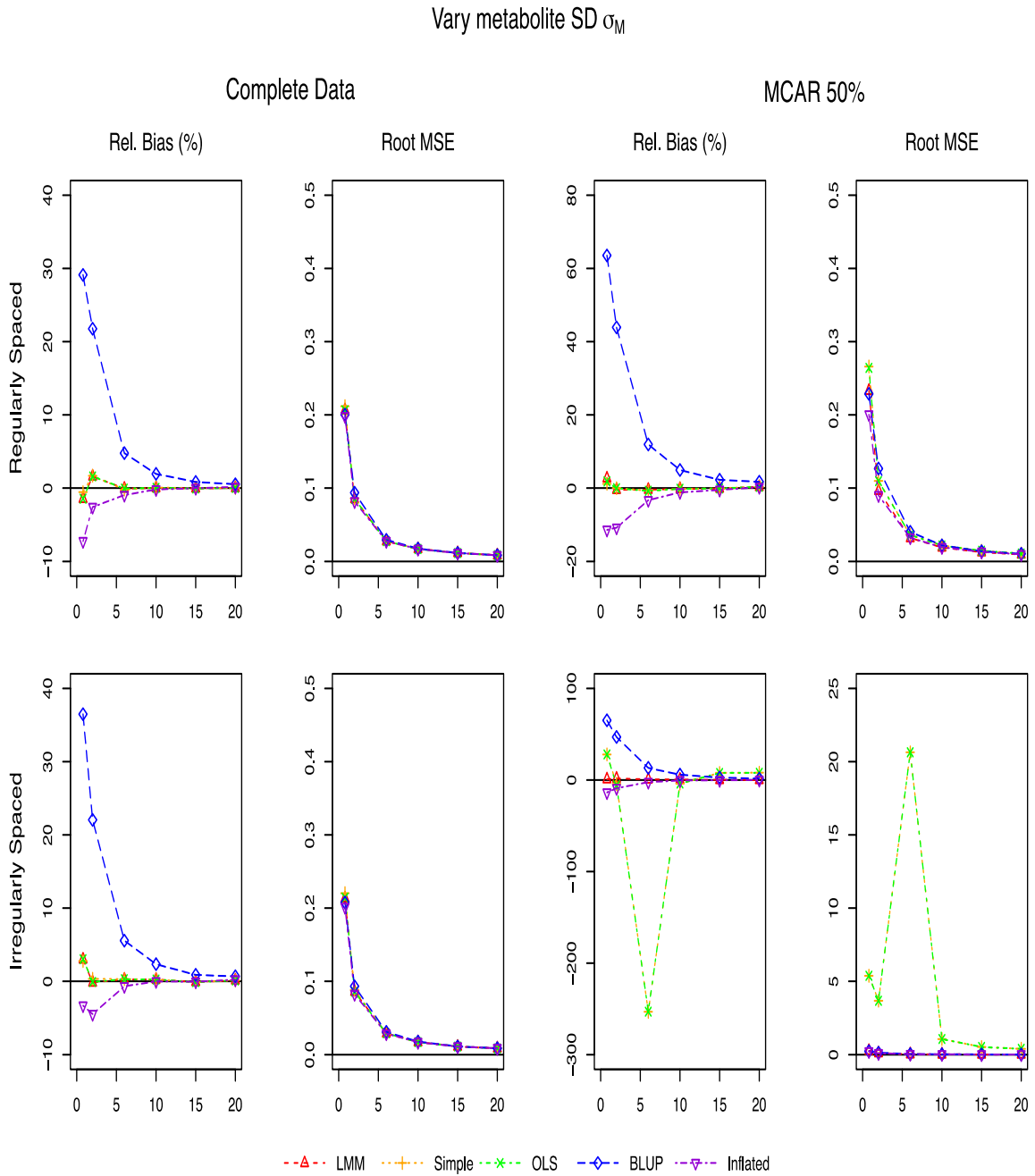

Main Figure 2 (Rel. Bias, Root MSE for vary random slope SD)
Performance in relative bias (%) and root MSE of our methods in estimating the association between annual rate of eGFR change and metabolite for the linear mixed model ($\hat{\beta}_3$) versus two-stage methods ($\hat{\alpha}_1$) as a function of random slope SD for the regularly and irregularly spaced cases of Complete Data and MCAR 50%.

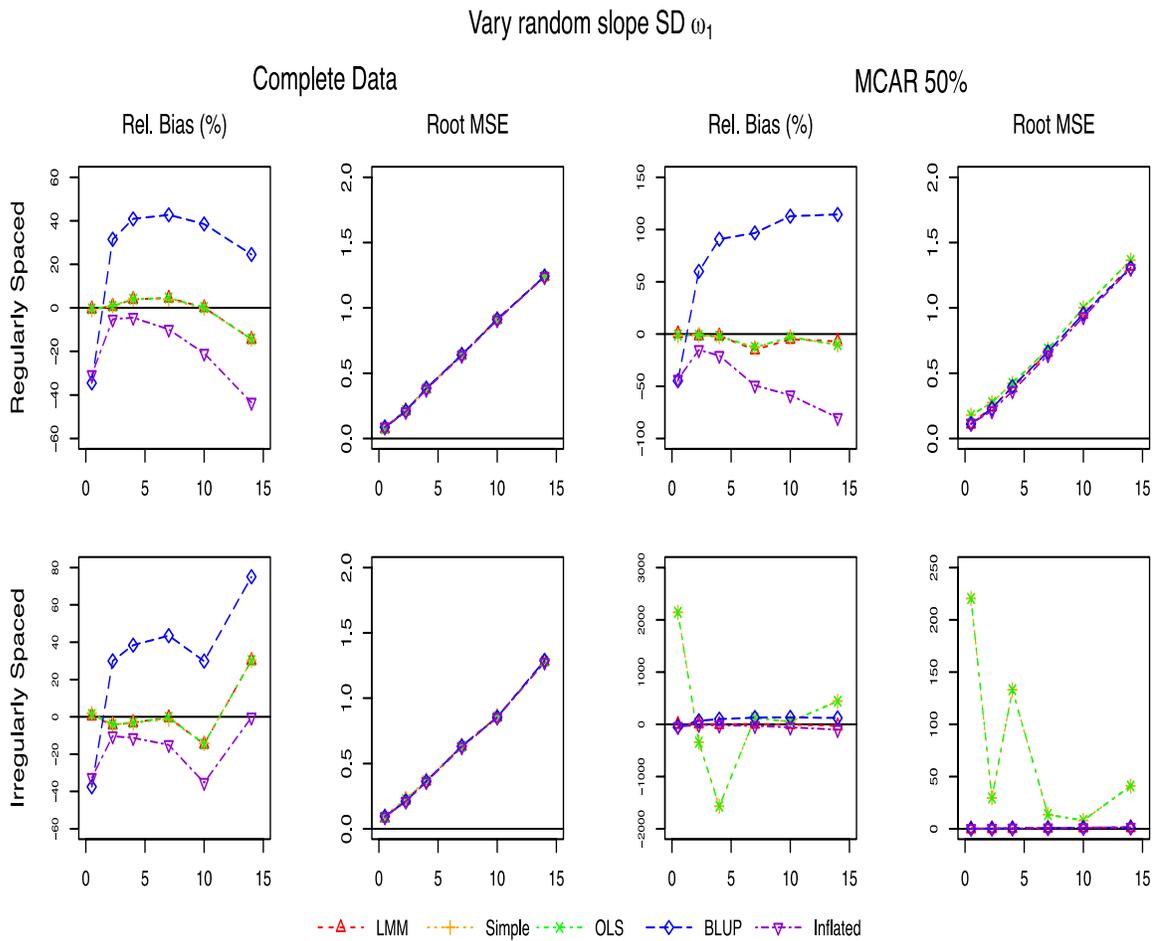

Main Figure 3 (Rel. Bias, Root MSE for vary random effects corr.)
Performance in relative bias (%) and root MSE of our methods in estimating the association between annual rate of eGFR change and metabolite for the linear mixed model ($\hat{\beta}_3$) versus two-stage methods ($\hat{\alpha}_1$) as a function of the correlation between random intercept and slope for the regularly and irregularly spaced cases of Complete Data and MCAR 50%.

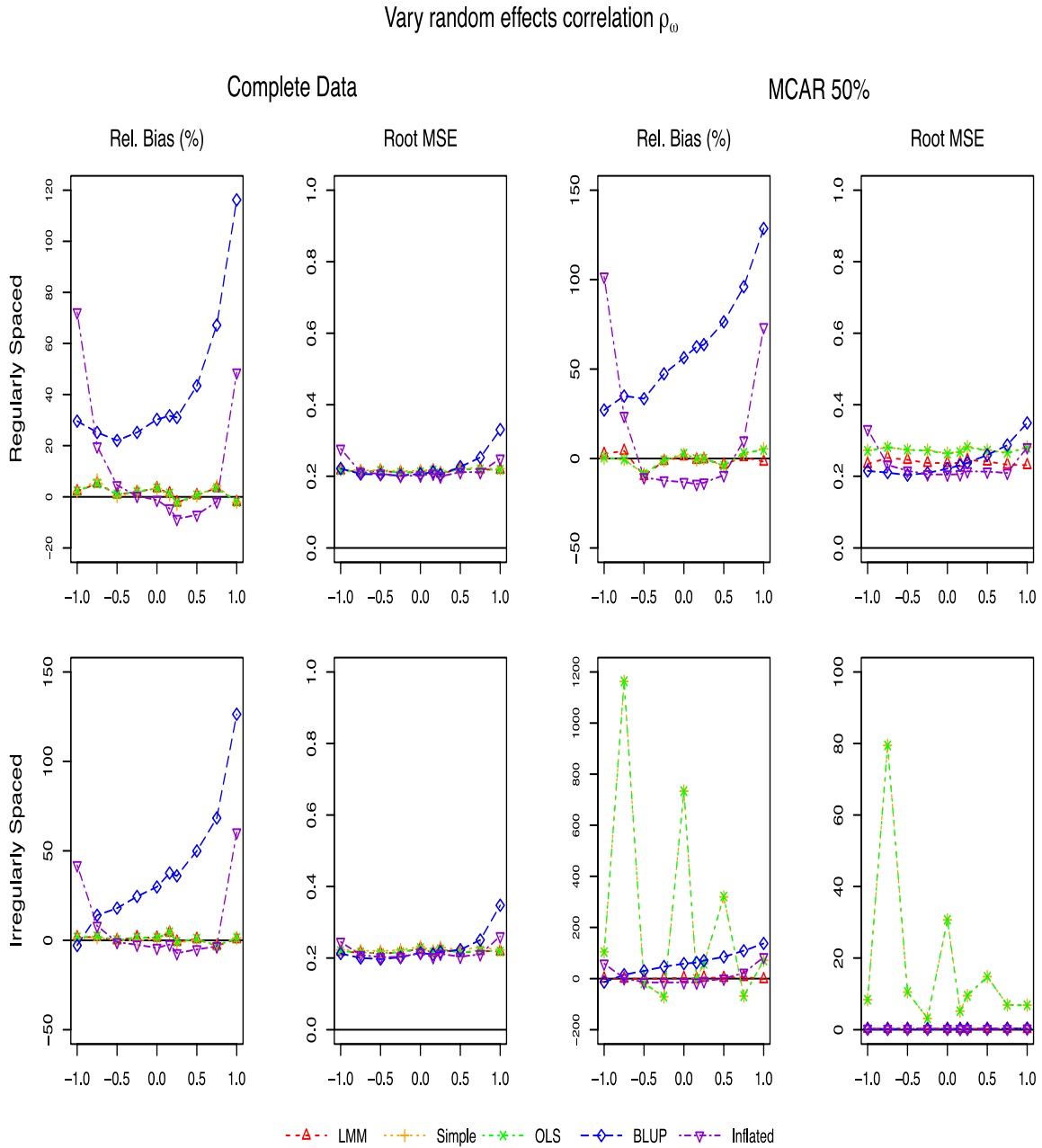

Supplementary Figure 1 (Bias, SD, SE for Main Figure 1)

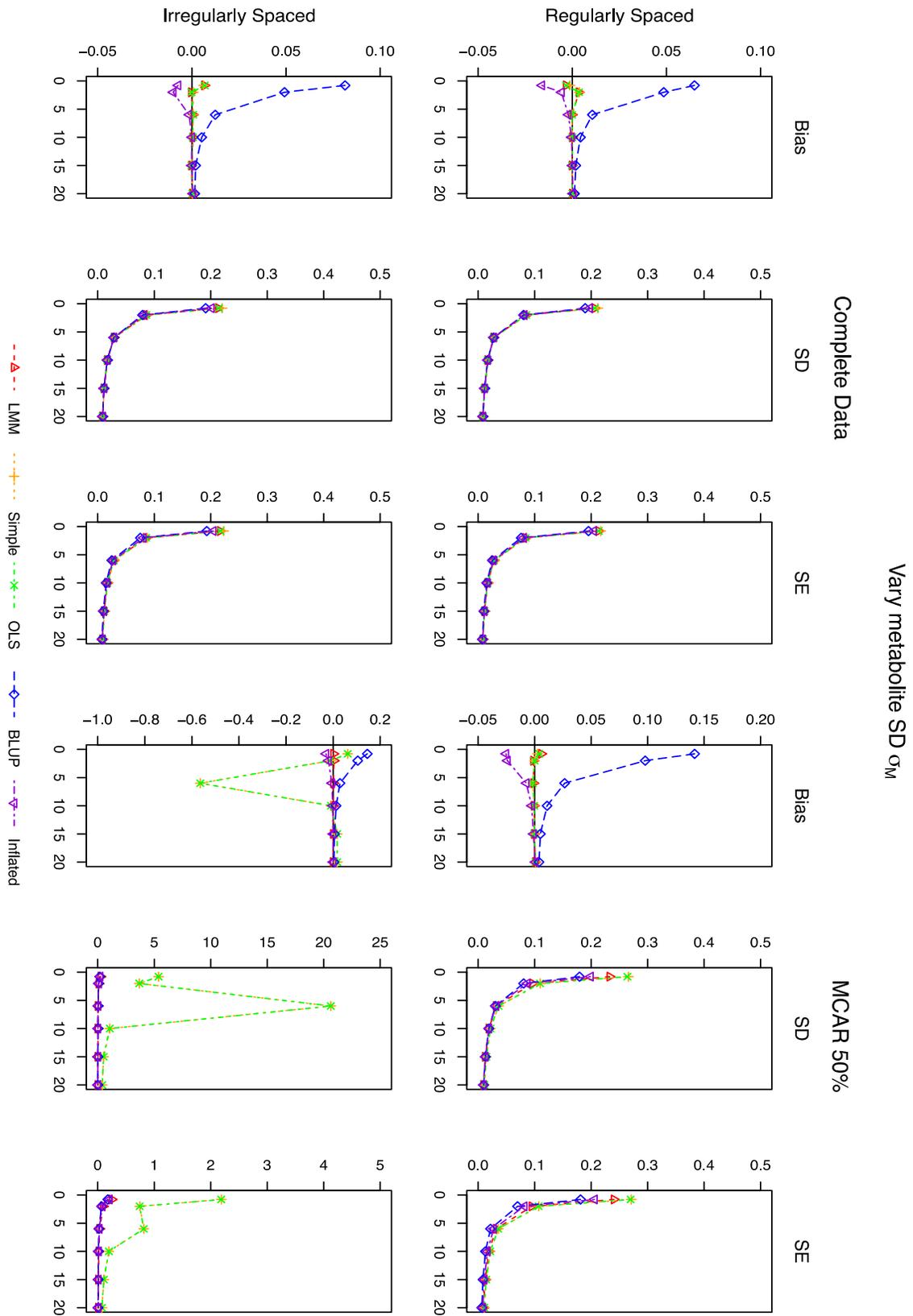

Supplementary Figure 2 (Bias, SD, SE for Main Figure 2)

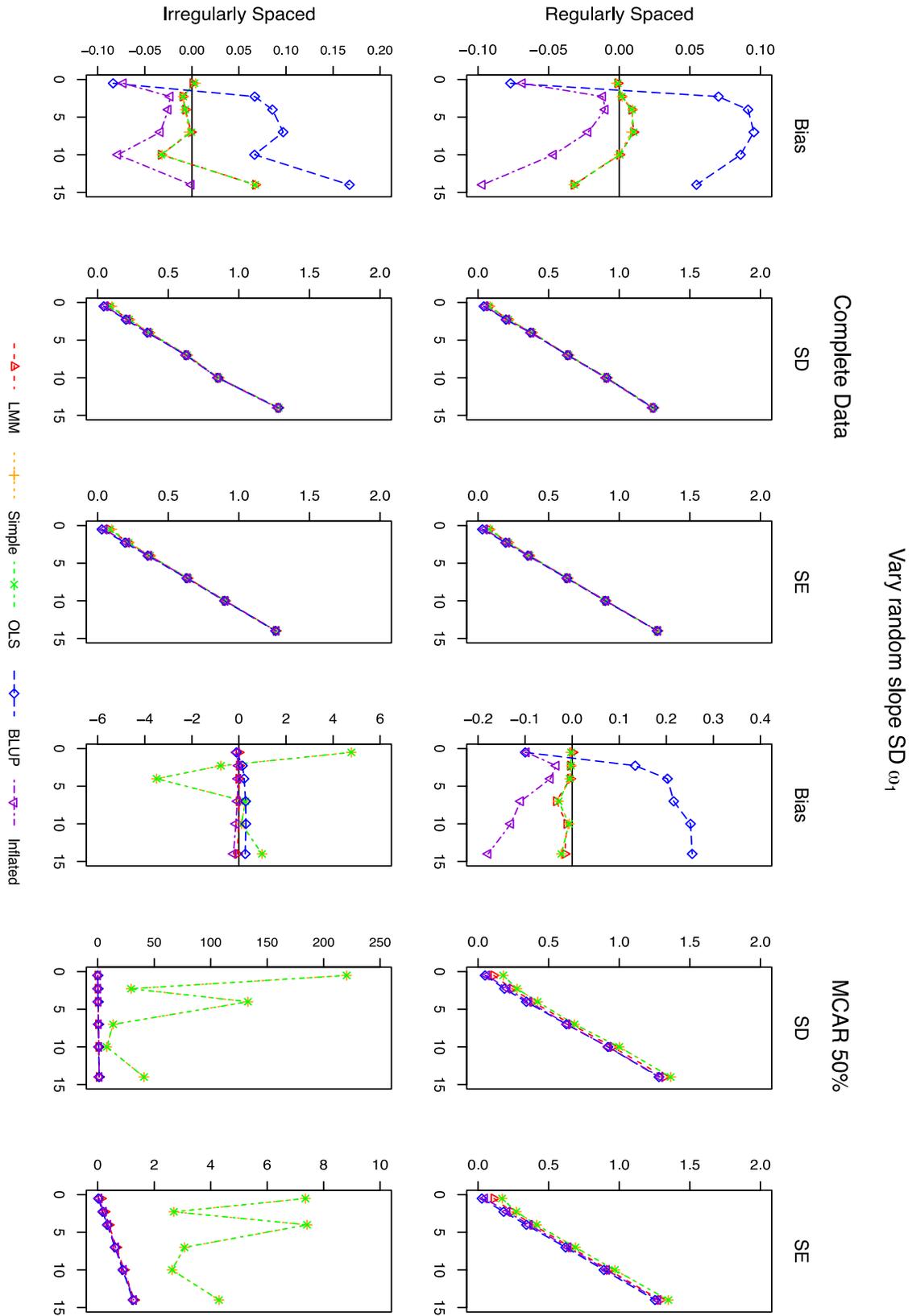

Supplementary Figure 3 (Bias, SD, SE for Main Figure 3)

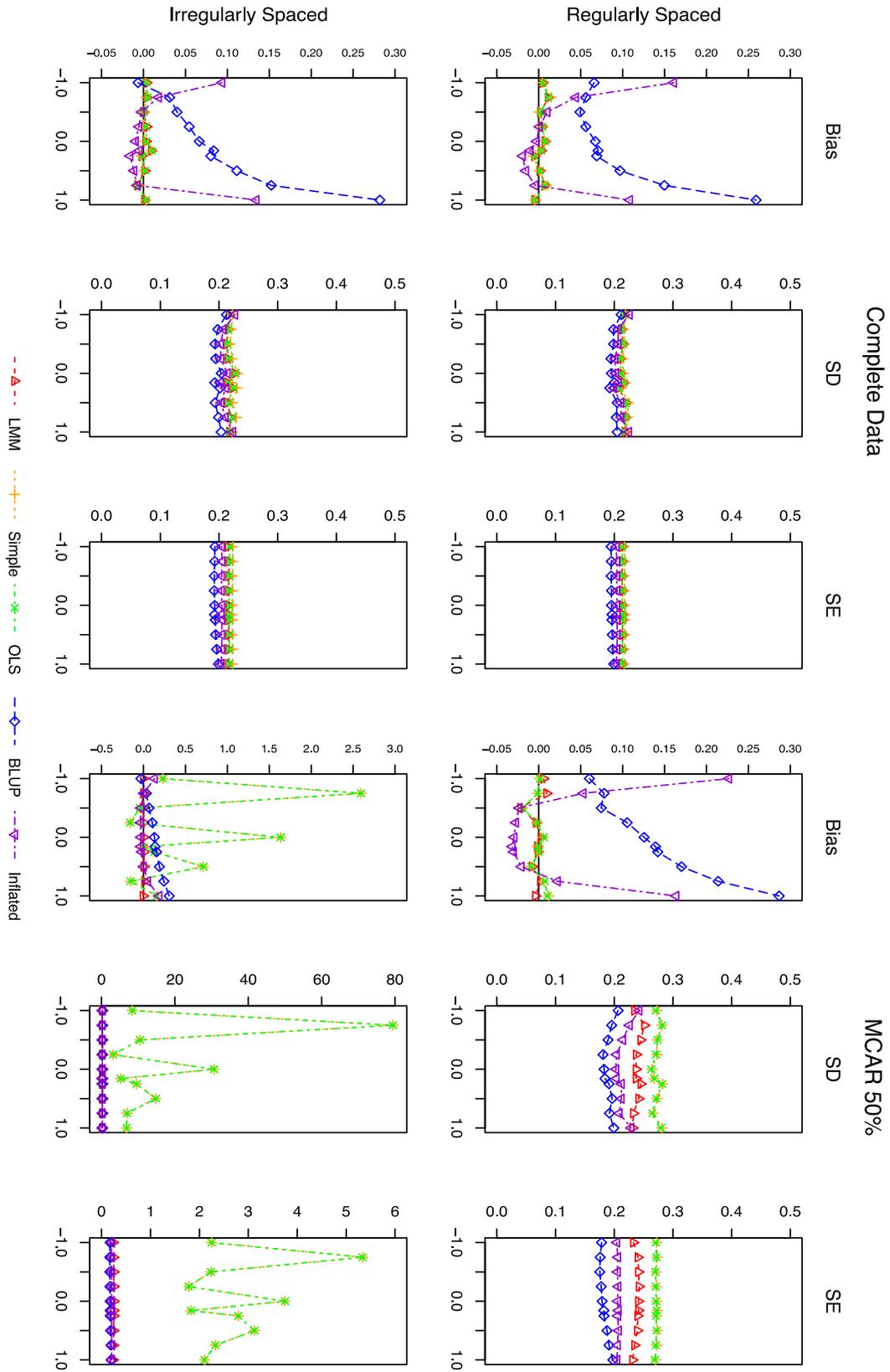

Supplementary Figure 4a (Rel. Bias, Root MSE for vary random intercept SD)
Performance in relative bias (%) and root MSE of our methods in estimating the association between annual rate of eGFR change and metabolite for the linear mixed model ($\hat{\beta}_3$) versus two-stage methods($\hat{\alpha}_1$) as a function of random intercept SD for the regularly and irregularly spaced cases of Complete Data and MCAR 50%.

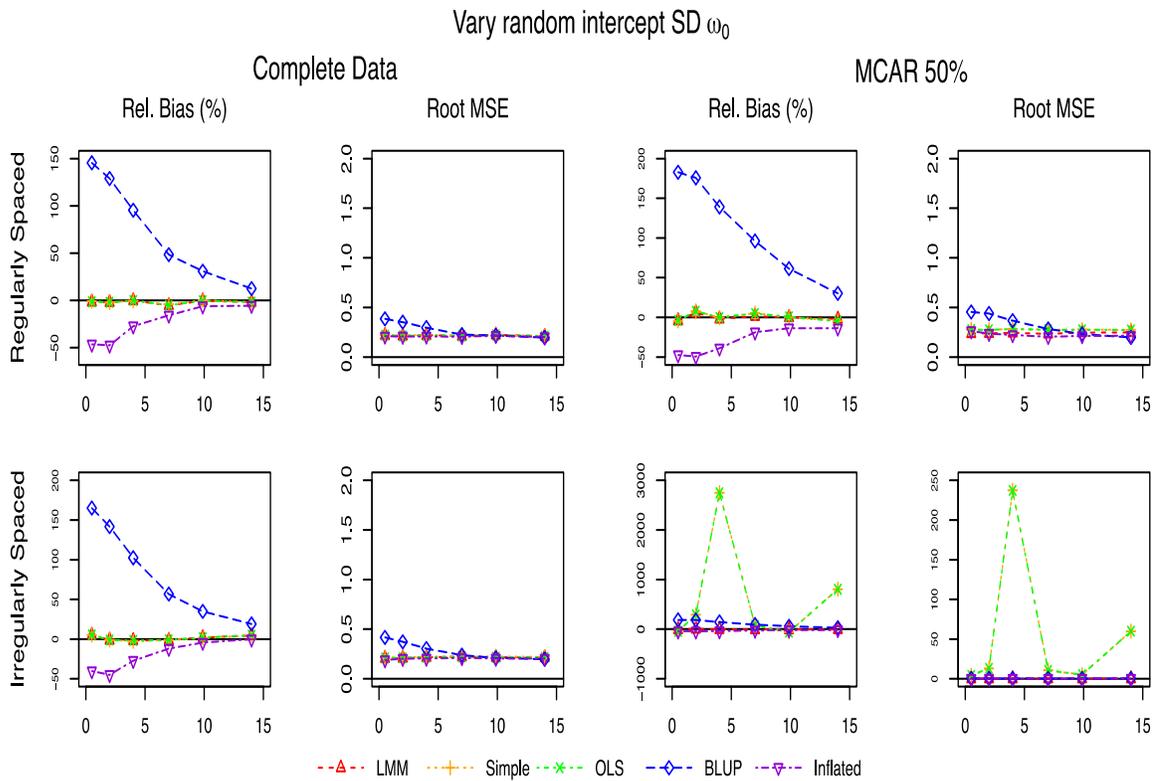

Supplementary Figure 4b (Bias, SD, SE for Supplementary Figure 4a)

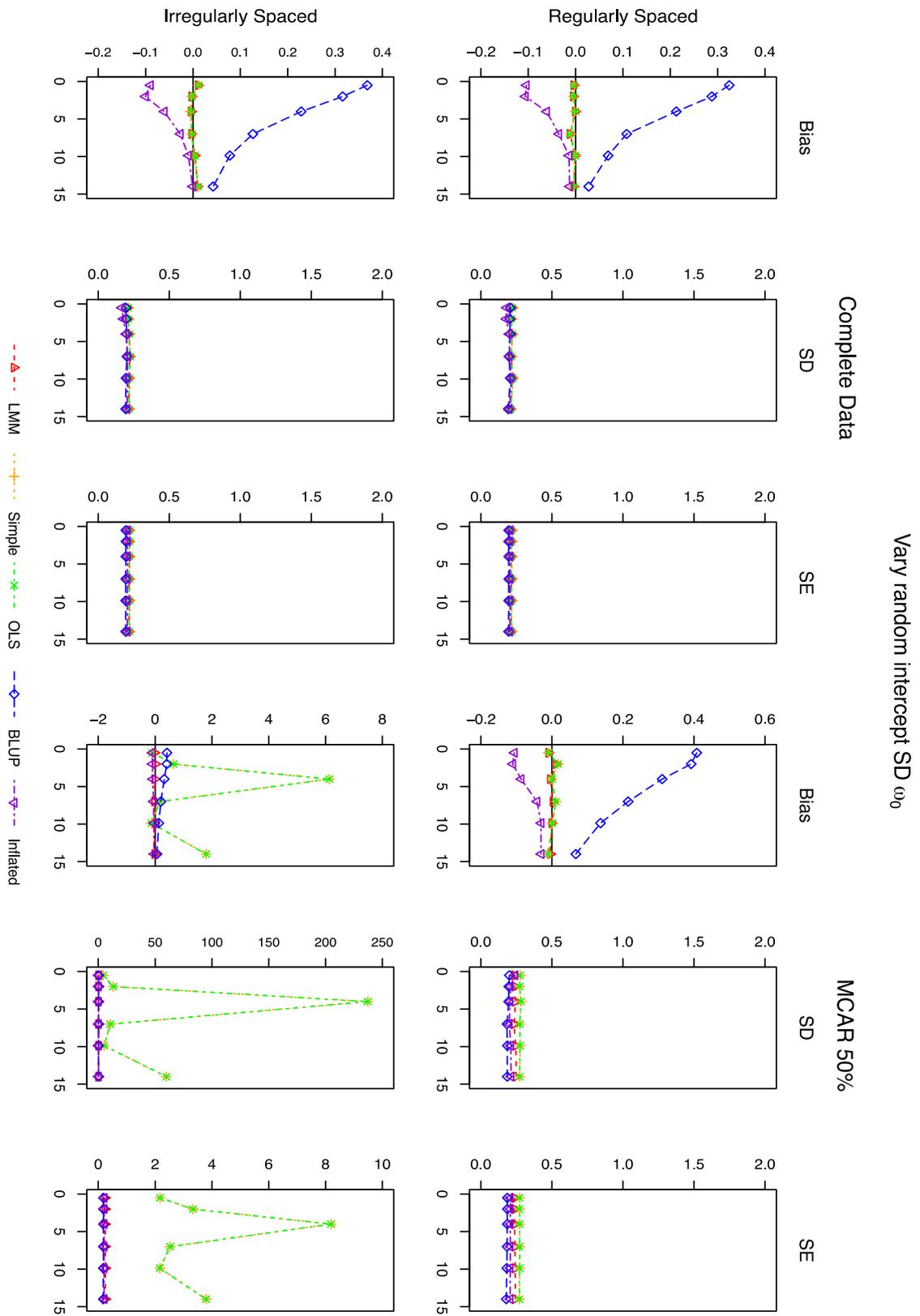

Supplementary Figure 5a (Rel. Bias, Root MSE for vary error SD)
Performance in relative bias (%) and root MSE of our methods in estimating the association between annual rate of eGFR change and metabolite for the linear mixed model ($\hat{\beta}_3$) versus two-stage methods ($\hat{\alpha}_1$) as a function of error SD for the regularly and irregularly spaced cases of Complete Data and MCAR 50%.

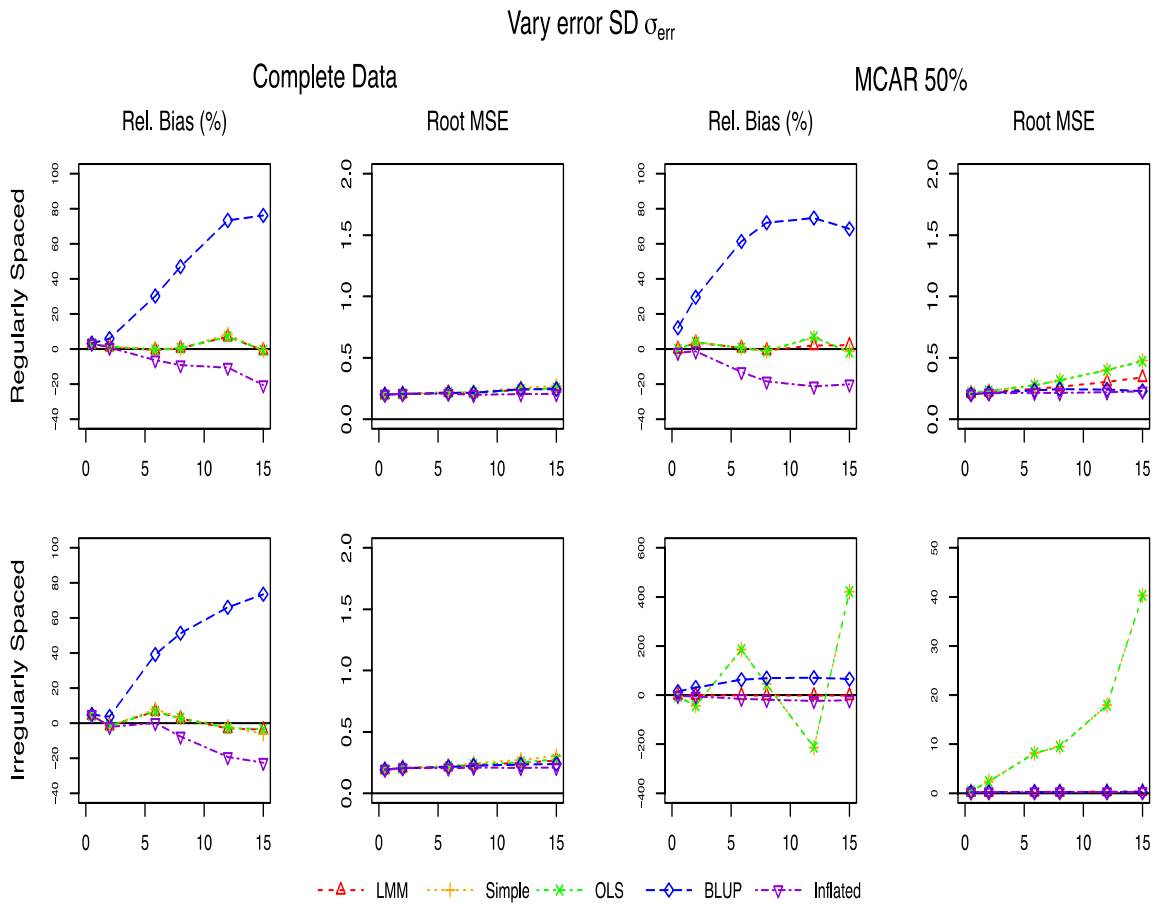

Supplementary Figure 5b (Bias, SD, SE for Supplementary Figure 5a)

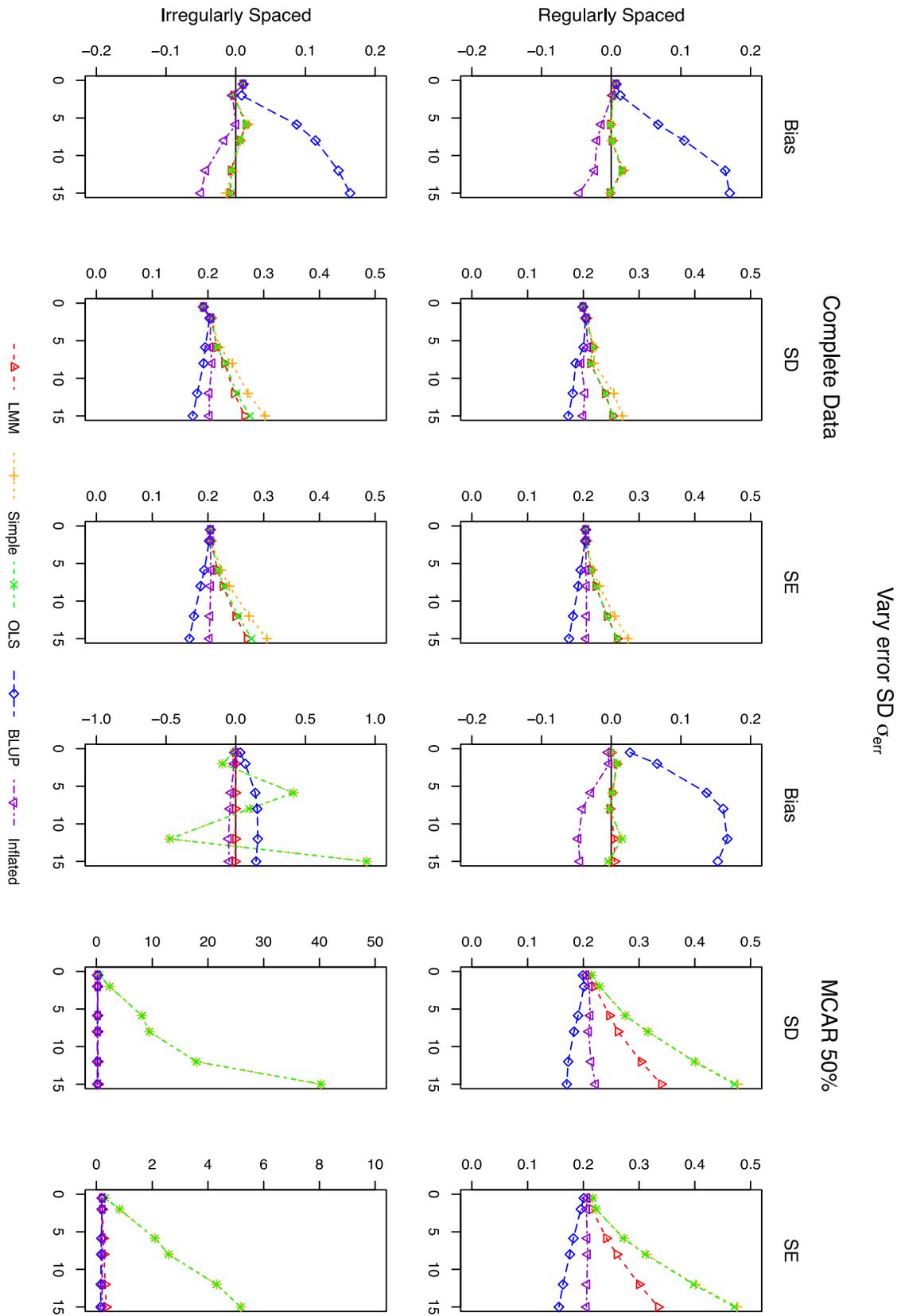